\documentclass[9pt,twocolumn,twoside]{opticajnl}
\journal{opticajournal} 
\setboolean{shortarticle}{false}

\usepackage{lineno}
\usepackage{siunitx}
\usepackage{hyperref}
\usepackage{comment}
\usepackage{graphicx}
\usepackage{amsmath}
\usepackage{amssymb}
\usepackage{bm}

\title{Online estimation of the transmission matrix of an atmospheric channel}
\author[1,*]{Giacomo Sorelli}
\author[1]{Douglas McDonald}
\author[1]{Szymon Gladysz}

\affil[1]{Fraunhofer IOSB, Ettlingen, Fraunhofer Institute of Optronics, System Technologies and Image Exploitation, Gutleuthausstr. 1, 76275 Ettlingen, Germany}

\affil[*]{giacomo.sorelli@iosb.fraunhofer.de}

\begin{abstract}  
We investigate the reconstruction of the transmission matrix of a time-evolving atmospheric channel with an online recursive optimization routine, using wave-optics simulations.
We demonstrate that this estimation technique is able to keep up with the evolution of the channel and enables a significant improvement of communication-relevant quantities such as the total power transmitted through the channel, and the coupling of the received light into a single mode fiber.
Moreover, we show that this approach is robust against measurement noise, and that it notably reduces the probability and duration of power outages, even in strong turbulence.
\end{abstract}

\begin{document}

\maketitle
\section{Introduction}
Atmospheric channels play an important role in various fields of science and technology including astronomy, remote sensing, imaging and communication \cite{Roddier04, Andrews05}.
All these rather diverse applications share the need of efficiently transmitting and/or decoding information encoded into electromagnetic radiation travelling through the channel.
Accordingly, it is crucial to identify strategies which efficiently mitigate the random wave distortions induced by the atmosphere.

For the past few decades, the main turbulence compensation strategy has been adaptive optics (AO) which uses a fast wavefront sensor working in closed loop with a wavefront shaping element (generally a deformable mirror) to correct  turbulence-induced phase distortions in real-time \cite{Roddier04, Tyson16}.
Compensating only phase fluctuations is sufficient for weak atmospheric layers, especially if located close to the receiver, i.e. in astronomy. 
On the other hand, AO fails to compensate for intensity fluctuations which become important in horizontal atmospheric channels, particularly under strong turbulence conditions \cite{Botygina_2019, Lukin_2020}, and can have a significant impact, e.g. in free-space quantum or classical communication \cite{Ren14, Sorelli19,Zhao20}.

In contrast to applications involving free-space propagation, in static scattering media wavefront shaping techniques have become the preferred tool to control light propagation \cite{Vellekoop:15}.
In this context, measuring the transmission matrix of the medium \cite{Popoff10} enables focusing \cite{Vellekoop07}, point-spread function engineering \cite{Boniface17}, high energy transmission (even through multiple scattering media) \cite{Kim12} and arbitrary field transmission \cite{Devaud22}.
Recent works applied these ideas to turbulence: efficient transmission of light through a specific static realization of an atmospheric channel was demonstrated in a table-top simulator \cite{Klug23}, and temporal evolution of the highly transmitting modes of an atmospheric channel was investigated in wave-optics simulations \cite{Bachmann23}.
These works demonstrated that efficient transmission of light can be achieved even through long horizontal channels under strong turbulence conditions; however, they both assumed perfect knowledge of the transmission matrix of the channel at a given instant in time.

The next essential step towards implementation of these promising light control techniques in real turbulence is the real-time acquisition of the transmission matrix of an atmospheric channel. 
This is a very challenging task as the typical time scale of atmospheric variations is of the order of a few milliseconds \cite{Roddier04}.
In this work, we investigate its feasibility by employing the recursive algorithm for online estimation of a time-dependent transmission matrix proposed in \cite{Valzania23}, while emulating realistic bandwidths of state-of-the-art wavefront shaping and sensing devices, such as those employed in high-end AO systems \cite{Gehner20, Anzuola16, Yu17, Abado10}.
We show, by extensive wave optics simulations, that this approach is suitable for the optimization of communication-relevant performance metrics, such as the total power collected by the receiver aperture and the total power coupled into a single-mode fiber, enabling a significant performance increase compared to transmission of standard mode bases \cite{Sorelli19, Cox19, Gu20}, especially in strong turbulence.

\section{Methods}
\label{sec:methods}
\subsection{Modeling a time-dependent atmospheric channel}
\begin{figure*}
    \centering
    \includegraphics[width = \textwidth]{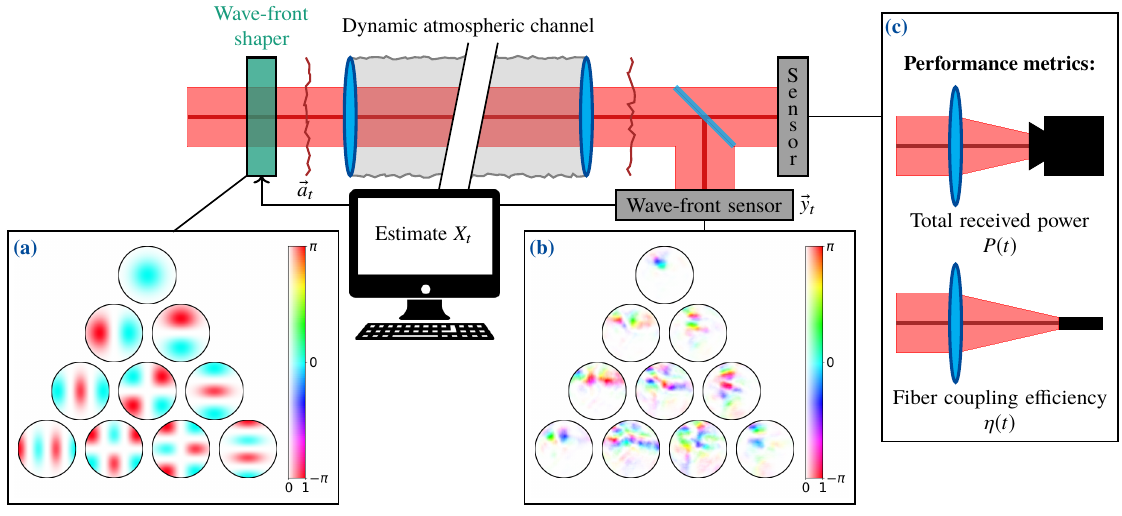}
    \caption{Schematic illustration of the setup for the online estimation of an atmospheric channel: a series of probe modes $a_t$ [here Hermite-Gaussian modes as illustrated in box (a)] are successively transmitted through the channel, the transmitted field $y_t$ is measured [see box (b)]. The input-output pairs $a_t$ and $y_t$ are used to update the estimation of the dynamic transmission matrix $X_t$, whose knowledge is used to compute the input field that optimizes a given performance metric, e.g. the total transmitted power or the power coupled into a single-mode fiber [see box (c)]. }
    \label{fig:scheme}
\end{figure*}
Propagation of a monochromatic wave $u(\bf{r})$ with wavelength $\lambda$ through an atmospheric channel is described by the {\it stochastic parabolic equation} \cite{Andrews05}
\begin{equation}
    - 2ik\,\frac{\partial u (\bf{r})}{\partial z} =  \nabla^2_\perp u({\bf r}) + 2k^2\,\delta n({\bf r}, t)\, u({\bf r}),
    \label{eq:stoch_par}
\end{equation}
where $k = 2\pi/\lambda$ is the wave number, $\nabla^2_\perp$ is the transverse (with respect to the propagation direction $z$) Laplace operator, and $\delta n({\bf r}, t)$ describes the turbulence-induced spatial and temporal fluctuations of the refractive index of the channel. 
The statistics of such random fluctuations are captured by the two-point correlation function, $\langle \delta n({\bf r}, t), \delta n({\bf r}^\prime, t) \rangle$, whose Fourier transform $\Phi_n(\kappa)$ is known as the {\it refractive index power spectrum}.
According to the Kolmogorov theory of turbulence \cite{Kolmogorov41a}, the three-dimensional refractive index power spectrum is given by
\begin{align}
    \Phi_n(\kappa) =  1.23\, C_n^2\, \kappa^{-11/3}, 
    \label{eq:kol_spec}
\end{align}
where $\kappa = |\bm {\kappa}|$, $\bm {\kappa}$ is the transverse angular wavenumber, and $C_n^2$ is the refractive index structure constant. 

For a wave propagating a distance $L$ through an atmospheric channel, the random spatial inhomogeneities of the refractive index described by \eqref{eq:kol_spec} induce phase distortions on the propagating light whose typical coherence length can be quantified (for a plane wave) by the {\it Fried parameter}: $r_0 = (0.42 \, k^2\, C_n^2\, L)^{-3/5}$ \cite{Andrews05}.
Upon propagation, phase distortions combine with diffraction resulting in {\it scintillation}, i.e. turbulence-induced intensity fluctuations, which are generally quantified through the {\it Rytov variance}: $\sigma_R^2 = 0.307 C_n^2 k^{7/6} L^{11/6}$ \cite{Andrews05}.

The refractive index structure constant $C_n^2$ (and consequently $r_0$ and $\sigma_R^2$) varies on typical time-scales of hours \cite{Andrews05} which are much longer than the typical duration of experiments. 
Accordingly, for propagation through horizontal channels (as those considered in this work), $C_n^2$ can be considered to be constant.
Nevertheless, an atmospheric channel evolves in time. According to {\it Taylor's hypothesis} \cite{Taylor35}, for short times ($t \lesssim 1$ s) such evolution can be described as a transverse flow of turbulent eddies due to the presence of wind. 
Under these assumptions, the coherence time of an atmospheric channel is fully defined by the ratio of its coherence length, i.e. the Fried parameter $r_0$, and the average transverse wind velocity $V$ across the channel: $t_c = 0.314\;r_0 / V$ \cite{Roddier04}.

An exact solution to \eqref{eq:stoch_par}, at a fixed time $t$, can be obtained numerically through the {\it split-step} method \cite{Martin88, Smith06, Lukin02}. 
This method consists of dividing the propagation path into segments (each short enough to ensure that intensity fluctuations are small) where all turbulence effects can be described as random phase screens connected by free diffraction in vacuum, with the latter easily implementable via Fourier optics methods \cite{Goodman05}. 
While at each propagation step only phase distortions are introduced, the combination of refraction on phase screens and diffraction allows to reproduce scintillation effects as well \cite{ Smith06, Lukin02}.
The accuracy of this technique relies on the generation of random phase screens that faithfully obey Kolmogorov statistics, which is possible following a variety of reliable techniques \cite{Martin88, Lane92, Johansson94, Roddier90, bachmann2024accurate}.   
Within this approach, the temporal evolution of the atmospheric channel can be simulated by transverse shifts of the phase screens according to a Gaussian velocity distribution with mean $V$ and variance $(\Delta V)^2$ \cite{Segel21}.
See Appendix A for further details.

\subsection{The transmission matrix and its online estimation}
Given the linearity of \eqref{eq:stoch_par}, the full propagation properties of an atmospheric channel can be described by a time-dependent transmission matrix $X_t \in \mathbb{C}^{M\times N}$ \cite{Rotter17,Carminati21}.
The matrix $X_t$ maps, at every time $t$, the coefficients $\vec{a}_t \in  \mathbb{C}^{N}$ of a field expansion in a set of transmitter modes $\left\lbrace \phi_j({\bm \rho}, 0)\right\rbrace_{j=0}^{N-1}$, into the coefficients $\vec{y}_t = X_t \vec{a}_t \in  \mathbb{C}^{M}$ of a field expansion in terms of the receiver modes $\left\lbrace \psi_j({\bm \rho}^\prime, L)\right\rbrace_{j=0}^{M-1}$, with ${\bm \rho}$ and ${\bm \rho}^\prime$ the transverse position vectors in the transmitter and receiver planes, respectively.
For an accurate representation of an atmospheric channel, Hermite-Gaussian (HG) modes (see box (a) in Fig.~\ref{fig:scheme}) can be chosen as a convenient transmitter basis $\left\lbrace \phi_j({\bm \rho}, 0)\right\rbrace_{j=0}^{N-1}$ as they are a solution to \eqref{eq:stoch_par} in the absence of turbulence, and therefore feature convenient diffraction properties \cite{Boucher21, Bachmann23}.
On the other hand, at the receiver side (see box (b) in Fig.~\ref{fig:scheme}), we represent the transmitted light in terms of the computationally and experimentally convenient pixel modes $\left\lbrace \psi_j({\bm \rho}^\prime, L)\right\rbrace_{j=0}^{M-1}$ \cite{Boucher21, Bachmann23}.

Knowledge of the transmission matrix $X_t$ can be used to engineer the optical field $u_t({\bm \rho}, 0)$ in the transmitter aperture, to optimize different properties of the field $u_t({\bm \rho}^\prime, L)$ in the receiver aperture. 
In particular, in this work we will consider two performance metrics highly relevant for communication experiments (see box (c) in Fig.~\ref{fig:scheme}): the total power collected by the receiver aperture $P(t)$, and the single-mode fiber coupling efficiency $\eta(t) = P_f(t)/ P(t)$ where $P_f(t)$ the total power coupled into the fiber \cite{buck_2004, winzer_1998}.
The total received power $P(t)$ can be optimized by performing a singular value decomposition of the transmission matrix $X_t = U_t \nu_t V_t^\dagger$, where $\dagger$ denotes the Hermitian transpose, and transmitting the input field $u_t({\bm \rho}, 0)$ which corresponds to the column of $V_t$ associated with the leading singular value in $\nu_t$ \cite{Rotter17, Miller19}.
To maximize the coupling efficiency $\eta(t)$, instead, we transmit the {\it phase conjugated} field $u_t({\bm \rho}, 0)$ corresponding to the coefficients $\vec{a}_t = X_t^\dagger \vec{y}_f$, where we introduced the coefficients $\vec{y}_f$ associated with the guided mode of the fiber, back-propagated to the aperture plane of the receiver \footnote{In our simulations, we consider a receiver which couples light to a standard single-mode fiber using a single coupling lens. Accordingly, the received field which maximizes the coupling efficiency can be determined by solving for the guided mode of the fibre, and numerically back-propagating this mode through the lens. The physical design of the lens is chosen to maximize coupling efficiency when the fundamental mode of the HG probe basis is propagated through the investigated channel ($L =1$ km) in the absence of turbulence. See Supplement 1 for further details.}.

To perform the above mentioned optimizations, at each time $t$, we need an estimate $\hat{X}_t$ of the transmission matrix $X_t$ of the atmospheric channel. 
As illustrated in Fig.~\ref{fig:scheme}, we can obtain such estimate using a recursive least-square (RLS) algorithm \cite{Valzania23}, which given the complex inputs $\vec{a}_t$ (prepared with a wavefront shaper) and the complex outputs $\vec{y}_t$ (acquired with a wavefront sensor), solves the optimization problem $\hat{X}_t = {\rm argmin}_{X_t} \mathcal{L}_t (X_t)$, with
\begin{align}
    \mathcal{L}_t (X_t) = \sum_{\tau = 1}^t \left(\lambda^{t-\tau}||\vec{y}_t - X_t \vec{a}_t||^2\right) + \delta \lambda^t ||X_t||_F^2
    \label{eq:LS_loss_function}
\end{align}
where $||\cdot||$ and $||\cdot||_F$ are the $L^2$-norm of a vector and the Frobenius norm of a matrix, respectively.
Equation \ref{eq:LS_loss_function} is a linear least square loss function, where, because of the {\it forgetting factor} $0 \ll \lambda < 1$, the {\it data fidelity terms} $||\vec{y}_t - X_t \vec{a}_t||^2$ are waited exponentially, so that measurements that occurred at $\tau \ll t$ are only marginally relevant to the present estimation.
This feature is crucial to keep up with the temporal evolution of the atmospheric channel.
On the other hand, the Tikhonov regularization term featuring the regularization constant $\delta$, biases the initial estimation of the transmission matrix.
Once $\lambda$ and $\delta$ are fixed, the linear least square problem has a unique solution that can be expressed in the form of normal equation \cite{Valzania23}
\begin{equation}
     C_t \hat{X}_t^\dagger = K_t,
     \label{eq:normal_eq}
\end{equation}
where at each time $t$ the inputs' covariance matrix $C_t$, and input-output cross covariance matrix $K_t$ are defined as
\begin{subequations}
\begin{align}
    C_t &= \sum_{\tau = 1}^t \left(\lambda^{t-\tau}  \vec{a}_t \vec{a}_t^\dagger \right) + \delta \lambda^t \mathbb{1}_N = \lambda C_{t-1} + \vec{a}_t \vec{a}_t^\dagger, \\
    K_t &= \sum_{\tau = 1}^t \left(\lambda^{t-\tau}  \vec{a}_t \vec{y}_t^\dagger \right) = \lambda K_{t-1} + \vec{a}_t \vec{y}_t^\dagger,
\end{align}
\label{eq:C_K}
\end{subequations}
with $\mathbb{1}_N$ denoting the $N$-dimensional identity matrix.
It is clear from \eqref{eq:C_K} that, at each time $t$, the least square estimator $\hat{X}_t$ of the transmission matrix can be constructed recursively by adding the new input and output data, $\vec{a}_t$ and $\vec{y}_t$ to the previous estimate of covariance and cross covariance matrices.
Accordingly, the RLS reconstruction algorithm is computationally very appealing, since it allows to consider the full history of the channel under investigation, while retaining in memory only information acquired during the previous iteration.
Finally, we point out that while direct inversion of \eqref{eq:normal_eq} is possible, $\hat{X}_t = K_t^\dagger(C_t^{-1})^\dagger$, it is generally preferable to use numerically stabler matrix inversion techniques, e.g. based on the {\it QR-}decomposition \cite{Alexander93}.

In principle, choosing large transmitter and receiver bases, i.e. large values of $N$ and $M$, allows for a more complete and accurate reconstruction of the transmission matrix $X_t$ \cite{Bachmann23}. 
However, as discussed in the algorithm presented above, the inputs $\vec{a}_t$ are transmitted sequentially. 
In this work we assume realistic bandwidths of both the wavefront shaping and sensing devices. 
With respect to shaping devices, modern micromirror arrays, which can be used to modulate both phase and amplitude, can achieve update rates of $3.6$ kHz at a spatial resolution of $512 \times 320$ pixels \cite{Gehner20}, while binary digital micromirror devices, which in principle could be used to perform the same task at en even higher spatial resolution of $1$ Mpix, can reach update rates of $20$ kHz, albeit with much lower diffraction efficiency (between 1 and 10 \%) \cite{Anzuola16, Yu17}. 
As for wavefront sensors, in this work we focus on Shack-Hartmann wavefront sensors which are capable of measurement rates as high as $100$ kHz \cite{Abado10}. 
In this work we consider slightly more modest device capabilities, specifically we assume a bandwidth for both the sensing and shaping devices of $r_{\rm mode} = 5$ kHz.
When we compare this with typical turbulence coherence times $t_c \sim 1$ ms \cite{Roddier04}, we see that the knowledge acquired from older measurements gets quickly outdated.  
Moreover, information from higher-order modes is often less relevant for the optimization of the performance metrics we are interested in. 
This is particularly true for the total transmitted power $P(t)$ as will be obvious from the results in Sec.~\ref{sec:results}.
To mitigate this effect, it is more convenient to repeatedly transmit a basis with fewer modes, such that the contribution from the most relevant modes is frequently updated, rather than transmit a basis with many (less relevant) modes.
In particular, in our simulations we used the $N=10$ HG modes as illustrated in Fig.~\ref{fig:scheme} (a). 
Accordingly, when the modes are transmitted sequentially, the full $N = 10$ modes basis is transmitted in $N/r_{\rm mode} = 2$ ms.
At receiver side instead we used a square grid with $M = 16\times 16 = 256$ pixels, commensurate with the resolution of standard wavefront sensors \cite{Abado10}. 

\section{Results}
\label{sec:results}
\subsection{Weak turbulence}
\label{sec:weak}
\begin{figure}[ht!]
    \centering
    \includegraphics[width = 0.48\textwidth]{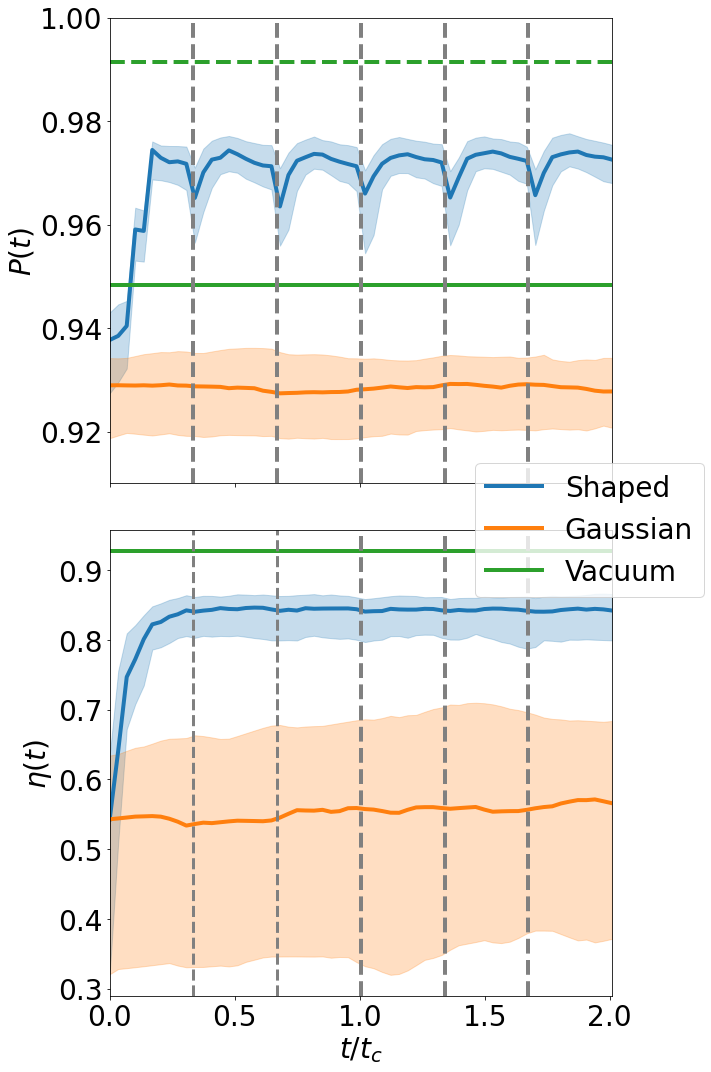}
    \caption{Total transmitted power $P(t)$ (top) and fiber coupling efficiency $\eta(t)$ (bottom) as a function of time $t$ (in units of the turbulence coherence time $t_c$). 
    Different colors correspond to: an input mode optimized using the current estimation of the transmission matrix $X_t$ transmitted in weak turbulence (blue), the fundamental mode of the HG probe basis (mode waist $w_0 = D/2\sqrt{2}$) transmitted in weak turbulence (orange) and in vacuum (green). 
    Solid lines corresponds to median, and shaded areas denote the first inter-quartile range computed from a sample of $500$ independent realizations of the atmospheric channel.
    The green dashed line in the top panel represents the total transmitted power in vacuum for a Gaussian beam with waist $w_0 = D/4$. 
    Vertical grey dashed lines denote the starting point of each basis repetition.}
    \label{fig:weak_scintillation}
\end{figure}
Using the techniques discussed in Sec.~\ref{sec:methods}, we simulate the estimation of the transmission matrix $X_t$ of a $L=1$ km long atmospheric channel delimited by transmitter and receiver apertures of equal diameter $D = 10$ cm, for a monochromatic beam of wavelength $\lambda  = 1550$ nm.
We start by studying this channel under weak turbulence conditions, in particular, we assume a refractive index structure constant $C_n^2 = 1.7 \times 10^{-14}$ m$^{-2/3}$, corresponding to Fried parameter $r_0 = 5$ cm, and Rytov variance $\sigma_R^2 = 0.337$.
Furthermore, we assume the transverse wind speed to have a Gaussian distribution with mean $V = 3$~m/s and variance $(\Delta V)^2 = 1\;{\rm m}^2/{\rm s}^2$, which results in the channel coherence time $t_c = 5.97$ ms.

In Fig.~\ref{fig:weak_scintillation}, we present the time evolution of the total transmitted power $P(t)$ and of the fiber coupling efficiency $\eta(t)$ optimized, as discussed in Sec.~\ref{sec:methods}, using the current estimation $\hat{X}_t$ of the transmission matrix of the atmospheric channel (blue solid lines with shaded inter-quartile region).
For comparison, we also report the same quantities for the fundamental HG mode with beam waist $w_0 = D/2\sqrt{2} = 35$ mm, transmitted both in vacuum (green solid lines) and through the atmospheric channel (orange solid lines with with shaded inter-quartile region). 

The first interesting thing to observe in the top panel of Fig.~\ref{fig:weak_scintillation} is that the total transmitted power $P(t)$ for the optimized mode does not only outperform the Gaussian mode $\phi_0({\bm \rho}, 0)$ in turbulence, but also in vacuum. 
This reveals, on the one hand, that the beam waist of  $\phi_0({\bm \rho}, 0)$ is not optimal for transmission in vacuum: a smaller beam with $w_0 = D/4 =25$ mm (dashed green line) would perform better. 
On the other hand, this shows how efficient the transmission optimization enabled by the online estimation of the transmission matrix is, despite it being completely agnostic to the properties of the atmospheric channel.
Another notable feature of the behavior of the total transmitted power $P(t)$ for the optimized modes is its slight decay at the end of each basis repetition (see dashed vertical lines in Fig~\ref{fig:weak_scintillation}). 
As anticipated in Sec.~\ref{sec:methods}, this is due to the fact that the largest contribution to the highest transmitting modes comes from lower order HG modes of the transmitter basis $\left\lbrace \phi_j({\bm \rho}, 0)\right\rbrace_{j=0}^{N-1}$, e.g. the fundamental mode $\phi_0({\bm \rho}, 0)$ already achieves a $~93\%$ transmissivity. 
Accordingly, while the contribution from several higher modes could further improve the transmission, as shown in \cite{Bachmann23}, such contribution cannot be acquired fast enough to keep up with the evolution of the channel. 

\begin{figure}[t!]
    \centering
    \includegraphics[width = 0.48\textwidth]{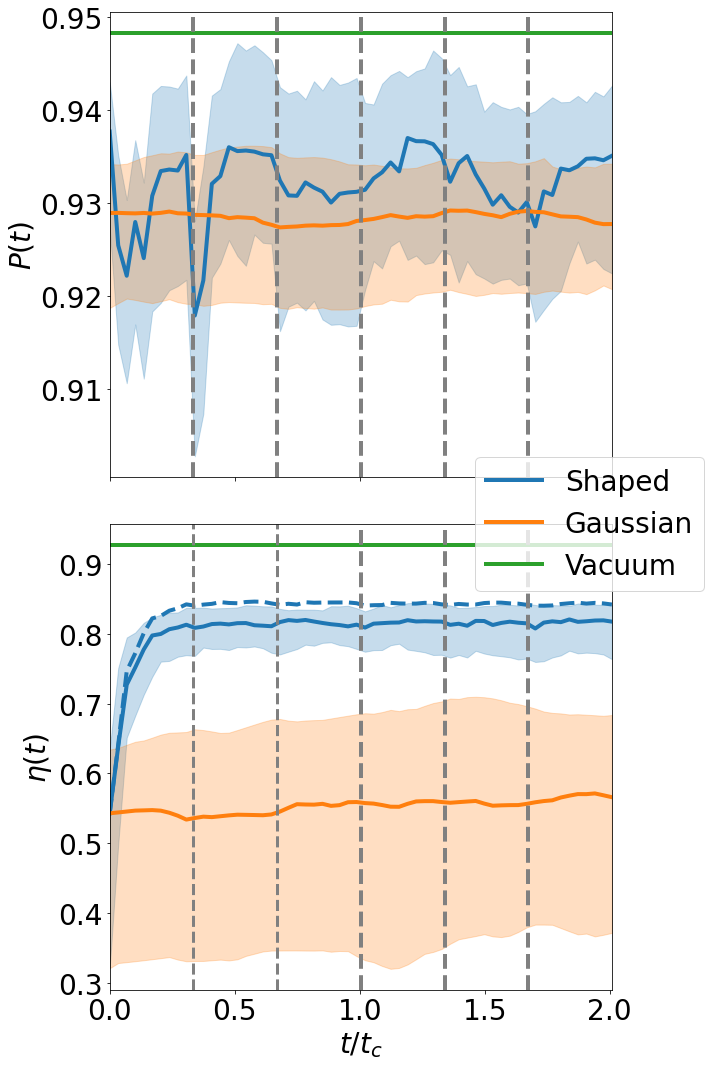}
    \caption{Total transmitted power $P(t)$ (top) and fiber coupling efficiency $\eta(t)$ (bottom) as a function of time $t$ (in units of the turbulence coherence time $t_c$). 
    Different colors correspond to: an input mode optimized using the current estimation of the transmission matrix $X_t$ (performed with noisy measurements) transmitted in weak turbulence (blue), the fundamental mode of the HG probe basis (mode waist $w_0 = D/2\sqrt{2}$) transmitted in weak turbulence (orange) and in vacuum (green). 
    Solid lines corresponds to median, and shaded areas denote the first interquartile range computed from a sample of $500$ independent realizations of the atmospheric channel.
    The blue dashed line in the bottom panel represents the fiber coupling efficiency $\eta(t)$ optimized using the noiseless estimation of the transmission, the same as the median plotted in the bottom 
    Vertical grey dashed lines denote the starting point of each basis repetition.}
    \label{fig:weak_scintillation_noisy}
\end{figure}
The situation is quite different for the optimization of the fiber coupling efficiency $\eta(t)$ (bottom panel of Fig.~\ref{fig:weak_scintillation}).
First, we observe that the impact of even weak turbulence on this metric is far more severe than for the total transmitted power $P(t)$: for a Gaussian beam we go from $\eta(t) > 0.9$ in vacuum (green line) to a broad negatively skewed coupling efficiency distribution with a median $\eta(t) \sim 0.55$ (orange line with error bands).
In comparison, the optimized $\eta(t)$ is not only much higher (on average above $80\%$) but also performs significantly more consistently among different realizations of the atmospheric channel: the variance of the optimized coupling efficiency is about an order of magnitude smaller than that of the non-optimized one.
Furthermore, we note that, as opposed to what we observed for the total transmitted power, after the first basis transmission, the optimized coupling efficiency remains approximately constant.  
Additionally, when doing this comparison we should also take into account the magnitude of the enhancement enabled by the knowledge of the transmission matrix is very different for the two metrics: few percent for $P(t)$ as opposed to almost $50\%$ ($\sim$ 3 dB) for $\eta(t)$.

The promising results observed in Fig.~\ref{fig:weak_scintillation} were obtained assuming noiseless measurements. 
However, in any real-life scenario noise in the wavefront sensor will affect the accuracy of our transmission matrix estimation with consequent impact on the performance metrics we are trying to optimize.
In particular, with typical wavefront sensors, under the investigated atmospheric conditions, while amplitude noise is generally negligible, significant errors can be present on the phase of the reconstructed field. 
To investigate the impact of such noisy phase reconstruction, we repeated the analysis presented above by adding to each pixel of the measured field a random phase following a Gaussian distribution with zero mean, and a variance $(\Delta \varphi)^2 = 1\; \rm{rad}^2$.
The results are presented in Fig.~\ref{fig:weak_scintillation_noisy}.

The first thing we note is that a noisy estimation of the transmission matrix $\hat{X}_t$ becomes almost useless for the optimization of the total transmitted power $P(t)$ (top panel of Fig.~\ref{fig:weak_scintillation_noisy}).
However, in such weak turbulence conditions, the transmissivity $P(t)$ of the unoptimized Gaussian mode $\phi_0({\bf r}, 0)$ is already fairly high ($\sim 93\%)$, improving upon which is very demanding, and probably unnecessary from a practical view-point. 
In Fig.~\ref{fig:weak_scintillation}, it can be seen how even with noiseless measurements we can achieve only a few percent improvement, and that information acquired with higher order modes struggles to keep up with the evolution of the channel. 
Therefore it is not surprising that noise has a significant impact on such a challenging task.
On the other hand, when we look at the results for the fiber coupling efficiency $\eta(t)$, where the transmission-matrix-based optimization gave more impactful results, we see that the enhancement obtained with noisy measurements (solid blue line in the bottom panel of Fig.~\ref{fig:weak_scintillation_noisy}) is only slightly worse than the one observed in the ideal scenario (dashed blue line in the bottom panel of Fig.~\ref{fig:weak_scintillation_noisy}).

\subsection{Strong turbulence}
\label{sec:strong}
We now want to study how the performance of the transmission-matrix estimation approach changes when we transition to the strong turbulence regime. 
We now consider an atmospheric channel with the same geometry and the same wind distribution as the one presented in Sec.~\ref{sec:results}~\ref{sec:weak}, but with a refractive index structure constant $C_n^2 = 1\times10^{-13}\;{\rm m}^{-2/3}$.
This results in a Fried parameter $r_0 = 2$ cm, Rytov variance $\sigma_R^2 = 1.98$ and turbulence coherence time $t_c = 2.06$ ms.
If we compare these parameters with those of the weak turbulence channel investigated in Sec.~\ref{sec:results}~\ref{sec:weak}, we see that phase distortions are now happening on a finer length scale (we move from $r_0$ of the order of the aperture diameter $D$ to $D/r_0 =5$), and intensity fluctuations become important ($\sigma_R^2 = 1$ being the conventional boundary between the weak and the strong scintillation regime \cite{Andrews05}).
Finally, we see that the increase in turbulence strength results in a much faster evolution of the channel. 
This is particularly relevant for our online estimation of the transmission matrix. 
In fact, while in the weak turbulence channel we were able to repeat transmission of our probing basis approximately three times within one turbulence coherence time, we are now in a condition where the total basis transmission time is on the order of the coherence time $N/r_{\rm modes} \sim t_c$.

\begin{figure}[t!]
    \centering
    \includegraphics[width = 0.48\textwidth]{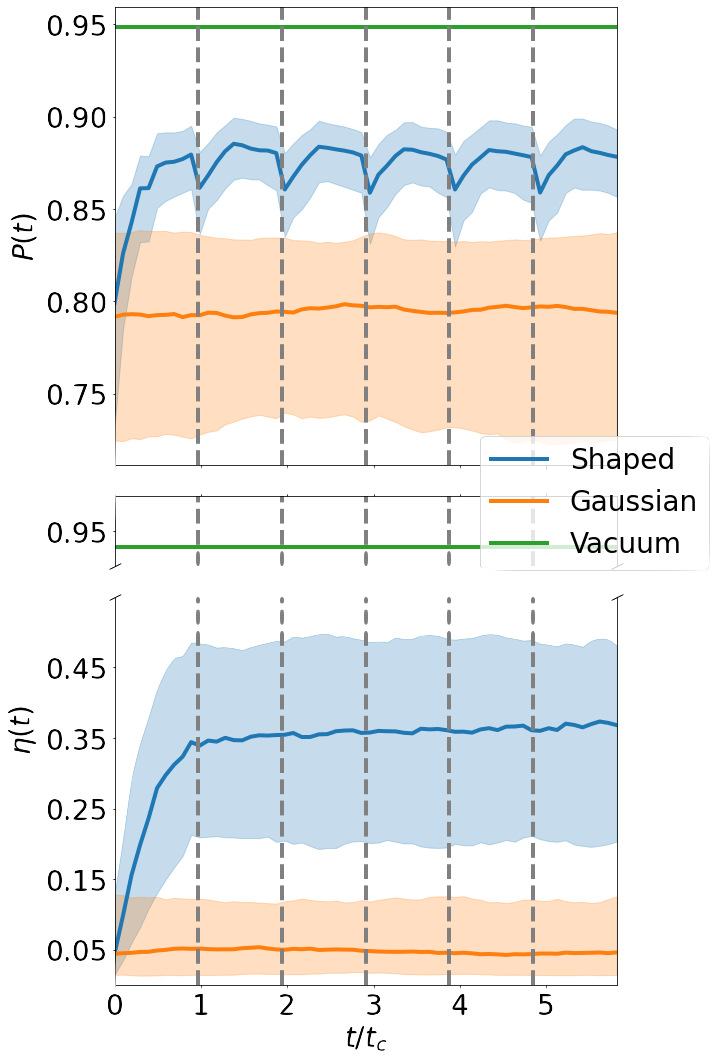}
    \caption{Total transmitted power $P(t)$ (top) and fiber coupling efficiency $\eta(t)$ (bottom) as a function of time $t$ (in units of the turbulence coherence time $t_c$). 
    Different colors correspond to: an input mode optimized using the current estimation of the transmission matrix $X_t$  transmitted in strong turbulence (blue), the fundamental mode of the HG probe basis (mode waist $w_0 = D/2\sqrt{2}$) transmitted in weak turbulence (orange) and in vacuum (green). 
    Solid lines corresponds to median, and shaded areas denote the first interquartile range computed from a sample of $250$ independent realizations of the atmospheric channel.
    Vertical grey dashed lines denote the starting point of each basis repetition.}
    \label{fig:strong_scintillation}
\end{figure}
In Fig.~\ref{fig:strong_scintillation}, we report the results of the optimization of the total transmitted power $P(t)$ and the fiber coupling efficiency $\eta(t)$ using, at each time $t$, the current estimation of the transmission matrix $\hat{X}_t$ under strong turbulence conditions.
The results are qualitatively similar to those observed in weak turbulence (Fig.~\ref{fig:weak_scintillation}), however the enhancement over the Gaussian beam $\phi_0({\bf r}, 0)$ achieved by the transmission-matrix-optimized modes is far larger in this case. 
This is particularly remarkable for the fiber coupling efficiency $\eta(t)$, where the relative improvement increases from $\eta_{\rm opt}(t)/\eta_{\rm Gauss}(t) \sim 1.5$ ($1.7$ dB) in weak turbulence to $\eta_{\rm opt}(t)/\eta_{\rm Gauss}(t) \sim 7$ ($8.4$ dB) in strong turbulence.

\begin{figure}
    \centering
    \includegraphics[width = 0.48\textwidth]{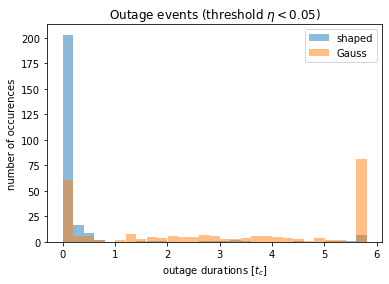}
    \caption{Histogram of the duration (in units of the turbulence coherence time $t_c$) of outage events, defined as intervals of time where less than $5\%$ of the output power is coupled into the single mode fiber, i.e. $\eta(t) < 0.05$. Results optimized using the current estimation of the transmission matrix $X_t$ (blue) are compared to those obtained transmitting the fundamental mode of the HG probe basis (orange).}
    \label{fig:outages}
\end{figure}
The enhancement in fiber coupling efficiency observed in the bottom panel of Fig.~\ref{fig:strong_scintillation} can make the difference between not being able to establish a communication link and efficiently maintaining it. 
In particular, in free-space optical communication it is common to establish a threshold value of the coupling efficiency $\eta_{\rm threshold}$ below which communication is deemed impossible, i.e. when $\eta(t) < \eta_{\rm threshold}$ we say that we are experiencing an {\it outage event} \cite{alexander_1997, Andrews05}.
Accordingly, when designing a free-space communication link it is important to assess the probability of outage events and the statistics of their duration.
We did this by analysing $250$ realizations of our strong turbulence channel considering a coupling efficiency threshold $\eta_{\rm threshold} = 0.05$, corresponding to $13$ dB fibre coupling loss compared to perfect coupling.
The results are presented in Fig.~\ref{fig:outages}. 
When we transmit the Gaussian mode $\phi_0({\bf r},0)$ (orange bars) the outage duration distribution is characterized by two almost equal peaks: one at $t=0$ representing {\it lucky} realizations where no outage occurs, and the other representing {\it unlucky} realizations when $\eta(t) < \eta_{\rm threshold}$ for the full duration of the simulation. 
In between those peaks we observe several events corresponding to outages whose duration is shorter than total simulation time.
On the other hand, when we shape the transmitted beam according to the current estimation of the transmission matrix $\hat{X}_t$, even though some unlucky realisations are still present, we see that about $80\%$ of the analysed channels fall into the lucky realizations peak.
Moreover, most of the outages of finite duration lasts less than one turbulence coherence time $t_c$. 
In fact, a closer look at the data reveals that all these outages occur during the first probe basis transmission (we recall $N/r_{\rm mode} \sim t_c$). 
In other words, we can experience outages while we are still learning the transmission matrix of the atmospheric channel, but once this knowledge is acquired, if we keep it up-to-date, then (excluding a very small number of unlucky realizations) we can keep $\eta(t) > \eta_{\rm threshold}$. 

\section{Conclusion}
We investigated the possibility of acquiring the transmission matrix of an atmospheric channel using a recursive online estimation technique \cite{Valzania23}, under both weak and strong turbulence conditions. 
We used such estimation to optimize communication-relevant performance metrics such as the total transmitted power $P(t)$ and the fiber coupling efficiency $\eta(t)$ under both weak and strong turbulence conditions.

Our results reveal that the optimization of the total transmitted power $P(t)$ is a very challenging task.
On the one hand, we know from \cite{Bachmann23} that acquiring a high-resolution estimation (e.g. using hundreds of probe modes) of the channel transmission matrix allows for the identification of modes with transmissivity close to unity even in strong turbulence. 
However, the same work revealed that the transmissivity of such modes decays exponentially fast when the atmosphere evolves in time. 
Accordingly, the $r_{\rm} = 5$ kHz update rate considered in this work (in accordance with state-of-the-art wavefront shaping and sensing technology) is not fast enough to reveal an enhancement that comes from several small contributions spread over many high-order modes.
These limitations are particularly evident in weak turbulence where the transmission matrix optimization enables only a modest enhancement which is completely obliterated by measurement noise. 
However, in this regime, a Gaussian mode already achieves a transmissivity above $90\%$, and optimization is arguably not necessary. 
On the other hand, in strong turbulence, despite these limitations, the transmission-matrix-based optimization enables an important enhancement of the transmitted power $P(t)$, and a significant reduction of its variability among different realizations of the atmospheric channel.

While the total transmitted power $P(t)$ is mostly affected by the transverse size of the transmitted waves and not by the finer details of their spatial distribution, that is not the case for the fiber coupling efficiency $\eta(t)$ which requires optimization of the spatial profile of the transmitted light in order to match that of the mode guided by the fiber.
Accordingly, this quantity is far more sensitive to turbulence-induced distortions.
Consequently, estimating, even imperfectly, the transmission matrix $\hat{X}_t$ of the atmospheric channel has a significant impact on the optimization of the fiber coupling efficiency $\eta(t)$.
This is clearly reflected in the robustness of this optimization technique to wavefront sensing noise we have observed in weak turbulence, as well as by its remarkable ability of almost completely eliminate the occurrence of outage events in strong turbulence.

Finally, in the present work we focused on horizontal channels, and on communication-relevant metrics. 
However, our approach can be easily adapted to consider slant or vertical paths and the estimated transmission matrix can be used for several different tasks such as focusing \cite{Vellekoop07, Popoff10} and point-spread function engineering \cite{Boniface17}.
This together with current technological advances in the development of wavefront sensors which are able to reliably operate at high-speed in strong turbulence conditions \cite{Zepp13,Zepp21,Zepp22}, paves the way to significantly enhance free-space communication and remote imaging both in ground-to-ground and ground-to-space applications.

\begin{backmatter}
\bmsection{Funding} This work was carried out during the tenure of an ERCIM ‘Alain Bensoussan’ Fellowship Programme.
This work was supported by the Fraunhofer Internal Programs under Grant No. Attract 40-09467.

\bmsection{Acknowledgments} GS is grateful to S. Gigan (LKB) and L. Valzania (Greenerwave) for their insights on the online estimation of the transmission matrix, and D. Bachman (Uni. Freiburg) for our discussion on wave optics simulations of atmospheric channels.

\bmsection{Disclosures} The authors declare no conflicts of interest.

\bmsection{Data Availability Statement} No data were generated or analyzed in the presented research.

\bmsection{Supplemental document}

\end{backmatter}
\bibliography{biblio}
\newpage

\appendix
\onecolumn
\section{Wave-optics simulations of atmospheric channels}
\label{sec:wave-optics}
We performed wave optics simulations on a $1024 \times 1024-$pixels numerical grid with the size of four times the diameter, $D = 10$ cm, of the transmitter and receiver apertures. 
To simulate free space propagation in vacuum, we used the angular spectrum propagator \cite{Goodman05, Schmidt10}. We generated random phase screens with Kolmogorov statistics (i.e. we did not include any inner or outer scales except for those enforced by the pixel and grid sizes in our numerical simulations) using the algorithm from \cite{Lane92} with $7$ subharmonics levels. 
To ensure the accuracy of our numerical simulations, we enforced a Rythov variance $\sigma_R^2 < 0.5$ for each partial propagation step. 
Accordingly, only one phase screen was  sufficient to simulate our weak turbulence channel, while seven screens where needed for our strong scintillation channel. 
All parameters of our wave optics simulations are summarized in Table~\ref{tab:wave-optics}.

\begin{table*}[htbp]
\centering
\caption{Parameters of the wave optics simulations}
\begin{tabular}{ccc}
{\bf Geometrical parameters} &   Weak turbulence & Strong turbulence\\
         \hline
         Input aperture  & \multicolumn{2}{c}{$D_{\rm in} = D = 10$ cm} \\
         Output aperture  & \multicolumn{2}{c}{$D_{\rm out} = D = 10$ cm} \\
         Channel length & \multicolumn{2}{c}{$L = 1$ km} \\
         {\bf Turbulence parameters} &   \\
         \hline
         Structure constant & $C_n^2 = 1.7\times 10^{-14}\; {\rm m}^{-2/3}$ & $C_n^2 =  10^{-13}\;{\rm m}^{-2/3}$\\
         Fried parameter & $r_0 = 0.05$ m & $r_0 = 0.02$ m \\
         Rytov variance & $\sigma_R^2 = 0.337$ (1 screen) & $\sigma_R^2 = 1.98$ (7 screens) \\
         Mean wind speed & \multicolumn{2}{c}{$\bar{V} = 3$ m/s} \\
         Wind speed variance & \multicolumn{2}{c}{$\Delta V = 1 m^2/s^2$}\\
         {\bf Mode parameters} &   \\
         \hline
         Input HG modes' waist & \multicolumn{2}{c}{$w_0 = D_{in}/2\sqrt{2} = 0.035$ m} \\
         Number of input modes & \multicolumn{2}{c}{HG modes with $n+m \leq 3$ ($N=10$ modes in total)} \\
         Output pixel modes size & \multicolumn{2}{c}{$d_{out} = 0.48$ mm} \\
         Output pixels number & \multicolumn{2}{c}{$M = 16 \times 16 = 256$} \\
         {\bf Fiber coupling} &   \\
         \hline
         Fiber Type & \multicolumn{2}{c}{Single-Mode: SMF-28 ULL \cite{corning_2014}} \\
         Optical wavelength & \multicolumn{2}{c}{$\lambda = 1550$ nm} \\
         Optimal mode waist (focal) & \multicolumn{2}{c}{$w_{\rm target} = 4.61$ \textmu m} \\
         Focal length & \multicolumn{2}{c}{$f = 0.35$ m} \\
\end{tabular}
  \label{tab:wave-optics}
\end{table*}

\section{Calculation of Fiber Coupling Efficiency}
\label{sec:fiber}

The degree to which an arbitrary signal can be coupled to an optical fiber is quantified by the coupling efficiency $\eta$, which is defined as the ratio of optical power coupled into the fiber $P_{c}$ to the available incident optical power in the plane of the fiber $P_{in}$. Ignoring losses due to Fresnel reflection at the fiber facet, the coupling efficiency for a single-mode fiber is given by the overlap integral \cite{buck_2004}
\begin{equation}
        \eta = \frac{P_{c}}{P_{in}} = \frac{\left| \iint_S u_{{in}}({\bm \rho}) \;  u^{*}_{0}({\bm \rho}) \, dS \right|^2}{ \iint_S \left| u_{{in}}({\bm \rho}) \right|^2 \, dS \; \iint_S \left| u_{0}({\bm \rho}) \right|^2 \, dS},
    \label{eq:coupling_efficiency}
\end{equation}
where $u_{in}({\bm \rho})$ and $u_{0}({\bm \rho})$ are the complex fields of the signal incident on the fiber and the guided mode of the fiber respectively, ${\bm \rho}$ is a transverse position vector located in the plane of the fiber, $^*$ denotes complex conjugation, and the integrals are evaluated over the entire (infinite) transverse plane at the fiber facet $S$. 

For a step-index optical fiber with a small refractive index constant (i.e.: weakly-guiding), the guided mode can be well described using the linearly polarized (LP) mode approximation and expressed as \cite{buck_2004, agrawal_2010}:
\begin{equation}
    u_0({\bm \rho}) = A_0
    \begin{cases} 
      J_0\left( \frac{p}{a} \left| \bm \rho \right| \right) & \left| \bm \rho \right| \leq a \\
      \frac{J_0\left( p \right)}{K_0\left( q \right)} K_0\left( \frac{q}{a} \left| \bm \rho \right| \right) & \left| \bm \rho \right| > a
    \end{cases}
    \label{eq:guided_mode}
\end{equation}
where $A_0$ is a scaling constant related to the power in the mode, $J_0\left( \cdot \right)$ is the Bessel function of the first kind, $K_0\left( \cdot \right)$ is the modified Bessel function of the second kind, $a$ is the radius of the fiber core, and $p$ and $q$ are dimensionless parameters related to the design of the fiber. In practice, the values of $p$ and $q$ are determined by numerically solving the fiber eigenvalue equation, given the core radius $a$, the fiber's numerical aperture $\textnormal{NA}$, and the wavelength of the signal $\lambda$. 

To compute the coupling efficiency for fields propagated through the atmospheric channels described in Sec.~\ref{sec:wave-optics}, we consider a simple optical system consisting of a single lens in the aperture plane which focuses the received beam to the fiber facet as shown in Figure 1 c) in the main text. The focal length of the lens used is chosen to optimize the coupling of the received Gaussian beam in the absence of turbulence. Specifically, a lens with focal length $f = 0.35$ cm is chosen such that the waist of the focused beam matches the waist of the guided fiber mode $\omega_0 = $ 4.61 \textmu m (which we estimate by approximating the fiber mode as a Gaussian beam \cite{marcuse_1977}) for a standard SMF-28 ULL fiber \cite{corning_2014} at $\lambda =$ 1550 nm. The same fiber is considered to numerically evaluate the expression of the guided mode given in Equation \ref{eq:guided_mode}. To avoid the computationally expensive operation of propagating each field from the aperture plane to the plane of the fiber, we rather back-propagate the guided mode through the coupling lens, which allows the integrals in Equation \ref{eq:coupling_efficiency} to be evaluated in the aperture plane \cite{winzer_1998}. The coupling efficiency is then computed numerically using the fields received in the aperture plane obtained via the wave optics simulations described in Sec.~\ref{sec:wave-optics}.

\end{document}